\documentclass[preprint,12pt,sort&compress]{elsarticle}
\usepackage{amsfonts}
\usepackage{amsmath}
\usepackage{amsthm}
\usepackage{amssymb}
\usepackage{graphicx}
\usepackage{dcolumn}
\usepackage{bm}
\usepackage{bbding}
\usepackage{mathrsfs}
\usepackage{graphicx}

\biboptions{numbers,sort&compress}

\begin{document}

\begin{frontmatter}
\title{Operational Resource Theory of Total Quantum Coherence}
\author{Si-ren Yang}
\author{Chang-shui Yu\corref{cor1}}
\ead{quaninformation@sina.com}
\cortext[cor1]{Corresponding author. }
\address{School of Physics, Dalian University of
Technology, Dalian 116024, China}
\begin{abstract}
Quantum coherence is an essential feature of quantum mechanics and is an important physical resource in quantum information. Recently, the resource theory of quantum coherence has been established parallel
with that of entanglement. In the resource theory, a resource can be well defined if given three ingredients: the free states, the resource, the (restricted) free operations.  In this paper, we study the resource theory of
coherence in a different light, that is, we consider the total coherence defined by the basis-free coherence maximized among all potential basis. We define the distillable total coherence and the total coherence cost and in
both the asymptotic regime and the single-copy regime show the reversible transformation between a state with certain total coherence and the state with the unit reference total coherence. Extensively, we demonstrate
that the total coherence can also be completely converted to the total correlation with the equal amount by the free operations. We also provide the alternative understanding of the total coherence, respectively, based on the entanglement and the total correlation in a different way.
\end{abstract}
\begin{keyword}
operational resource theory \sep total quantum coherence\sep transformation between states
\end{keyword}

\end{frontmatter}

\section{Introduction}

Quantum coherence is the essence of the interference phenomena and is the most
fundamental feature in quantum mechanics.
It is closely related to almost all the intriguing
quantum phenomena especially including the most notable  quantum entanglement \cite{entan1} and quantum discord \cite{9,10}.
Quantum coherence has been found to be an important physical resource that can be exploited to achieve various tasks in  many areas such as
 quantum biology \cite{108,hue1,reb,llo,abb,licm}, quantum thermal engine \cite{106,107,615,618,Hardral}, quantum transport \cite{trans1,trans2} and so on.
 Thus similar to the quantitative theory of entanglement \cite{horodec,virma,vedra,plenioeq} and other nonclassical resources \cite{adesso2016,horodec2013b,modi,sperling,strel}, a mathematically rigorous fashion (i.e., the resource theory of coherence) needs to be established for quantifying the resource character of coherence.
 Recently,  the resource theory of coherence has been proposed in Ref. \cite{aberg,plenio} and the relevant researches
 have attracted the increasing interest \cite{levi,yu2,yuan,Xi,Hu,zhangl,916,925,927,928,930,932,933,shengw,949,bagan,jha,radha,Killoran,GuangCanGuo,Alexej,Pati1,Fan1,Fan2,Girolami,Pati2,YuGuo,skewin}. In particular, Ref. \cite{plenio} pointed out that a good
 coherence measure should satisfy three properties: (i) the  incoherent states have no coherence; (ii) the average coherence isn't increased under the incoherent operations; (iii) the coherence is convex under the mixing
 of density matrices. Ref. \cite{strel}  showed the coherence in a state can be completely converted to the entanglement  with the equal amount and Ref. \cite{vedral} pointed out that the coherence can also be converted to the quantum discord. The operational
 resource theory of coherence was systematically described in Ref. \cite{winter} which showed the optimal conversion rates  between the states with different coherence in the asymptotic regime.

In fact, the resource theory is developed far beyond entanglement and coherence. A resource can be well defined so long as the three ingredients: the free states, the resource and the (restricted) free operations are
 clearly characterized.  For example, Ref. \cite{gour1} studied the resource theory of coherence by addressing different sets of incoherent operations, such as the maximal incoherent operations (IO), the strictly IO and
 the dephasing covariant IO, etc. A general discussion of the resource theory was presented in Ref. \cite{gour2}. One can note that all the mentioned coherence are the basis-dependent, which is consistent with our usual
 understanding of the coherence. This means that the coherence of a given state could be different under different basis. So a natural question is what the maximal coherence is in a state with all potential basis taken into
 account. This leads us to define the total coherence in a state \cite{yu}, which is obviously basis-independent. From the resource theory point of view, this actually redefines the free operations as done in Ref. \cite{gour1}. Physically, this treatment is also quite practical because the change of basis (e.g. rotating the plate for the photon state) isn't a problem at all in experiment.

 In this paper, we consider the total coherence as a resource and establish its resource theory paralleling with the resource theory of the basis-dependent coherence \cite{winter}. We define the distillable total coherence and the total
 coherence cost, revealing the transformational relation between the state with the resource (total coherence) and the state with the unit reference total coherence in both the single-copy regime and the asymptotic regime.
 In addition, we also show that the total coherence can be completely converted to the equal amount of the total correlation by the defined the incoherent operations assisted by the incoherent state. Finally, we give the alternative operational meanings based on the pure-state entanglement and the total correlation.
 This paper is organized  as follows. We first briefly review the properties of the total coherence, present the clear definitions of the distillable total coherence and the total coherence cost and show
 the transformational relation between the state with the certain total coherence and the state with unit reference total coherence.  Then we proved that the total coherence can be completely converted to the equal
 amount of total correlation and discuss some potential link with entanglement. Finally we present the operational meanings of the total coherence and summarize the paper.

\section{Operational resource theory for the total coherence}

To begin with, we review the total quantum coherence as well as its property.
The total coherence as mentioned previously is defined as the maximal coherence by optimizing
among all potential basis.
In Ref. \cite{yu} we have presented several good measures of the total coherence. However,
in this paper we would like to employ the total coherence measure in terms of the relative entropy  \cite{yu}. That is,
the total quantum coherence of an $n$ -dimensional quantum state $\rho$ is quantified by
\begin{equation}
C_{R}(\rho )=\mathrm{log}n-S(\rho
),\label{1}\end{equation}
where $S(\rho)=-\mathrm{Tr}\rho \mathrm{log}\rho$ is the von Neumann entropy of $\rho$.

\textit{Within the framework of the total coherence}, we have shown that $C_{R}(\rho )$ has the following properties \cite{yu}.

(1) The maximally mixed state $\rho_I=\frac{\mathbb{I}_{n}}{n}$ is the unique incoherent state, i.e., $C_{R}(\rho_I)=0$.

(2) The average total coherence doesn't change under the selective incoherent operation defined in the Kraus representation as $\$_I=\{U_i:\sum_ip_iU^\dag_i U_i=\mathbb{I}_n\}$ where $p_i$ denotes the probability
corresponding to the unitary operator $U_i$.

(2') The total coherence of the post-operation state doesn't increase under the incoherent operation $\$_I$. In particular, for any $\rho$ one can find an incoherent operation $\tilde{\$}_I$  leading $\rho$ to complete
decoherence, i.e,  $\tilde{\$}_I(\rho)=\rho_I$.

(3) $C_R$ is convex under the classical mixture of density matrices.

(4) All the pure states in the same state space have the same maximal total coherence.

 It is obvious that the state $\rho_I=\frac{\mathbb{I}_{n}}{n}$ is the free state, the operation $\$_I$ defines the (restricted) free operations. From all the properties one can find that the purity acts as the resource. In
 addition, one can find that (2) corresponds to the strong monotonicity and (2') corresponds to the monotonicity compared with the resource theory of the basis-dependent coherence. It is implied that the relative entropy
 given in Eq. (\ref{1}) is a good measure of the total coherence. In addition, based on the property of the logarithm and the von Neumann entropy, one can easily find that the above total coherence measure is additive, that
 is,

 (5) $C_R(\rho\otimes\sigma)=C_R(\rho)+ C_R(\sigma)$ for any two states $\rho$ and $\sigma$.

With these good properties, we can develop an operational
resource theory for the total coherence. We would like to begin with the transformation between the states with only a single copy as follows.

\textbf{Theorem 1}.-For the two \textit{d}-dimensional states $\rho$ and $\sigma$,  if $\rho$ can be transformed to $\sigma$ by the incoherent operation ${\$}_I$, then  $C_R(\rho)\geq C_R(\sigma)$.

\textbf{Proof}. It has been shown  in Ref. \cite{chef}  that for the two
\textit{d}-dimensional states $\rho$ and $\sigma$, $\rho$ can be transformed to
$\sigma$ by the incoherent operation ${\$}_I$ if and only if $\rho\succ\sigma$. 
Due to the concavity of $S(\rho)$, we have $S(\rho)\leq S(\sigma)$, that is,  $C_R(\rho)\geq C_R(\sigma)$ according to Eq. (\ref{1}).  To sum up, one can find our theorem is valid.\hfill $\blacksquare$
  
In our theoretic framework, one can find that all the pure states in the same state space have the same maximal coherence. Here we fix the qubit pure states as the unit reference, and then consider the optimal rate of the
asymptotic transformations between any given state and the unit reference (qubit pure states) by utilizing our incoherent operation $\$_I$. The transformation process can be summarized as follows \cite{horo}.

(\textit{Distillable total coherence}.-)Consider the asymptotic transformation from the $n$ copies of the state $\rho$ to the $m$ copies of some given qubit state $\left\vert\psi\right\rangle\left\langle\psi\right\vert$.
However, such a transformation is not generally exact even if $n$ tends to infinity. But it is possible to obtain a state $\sigma_n(\left\vert\psi\right\rangle\left\langle\psi\right\vert)$ which will asymptotically converge
to $\left\vert\psi\right\rangle\left\langle\psi\right\vert^{\otimes m}$. So the process can be formally formulated as
\begin{equation}
\rho^{\otimes n}\overset{\$_I}{\rightarrow}\sigma_n (\left\vert\psi\right\rangle\left\langle\psi\right\vert)\approx\left\vert\psi\right\rangle\left\langle\psi\right\vert^{\otimes m}
\end{equation}
such that
\begin{equation}
\lim_{n\rightarrow \infty}\left\Vert \sigma_n(\left\vert\psi\right\rangle\left\langle\psi\right\vert)-\left\vert\psi\right\rangle\left\langle\psi\right\vert^{\otimes m}\right\Vert_{tr}=0\label{condi}
\end{equation}
with $\left\Vert\cdot\right\Vert_{tr}$ denoting the trace norm. Since this process describes how much unit total coherence (qubit-pure-state coherence) can be asymptotically extracted from give states, the distillable total
coherence $C_{R}^{D}(\rho)$ similar to the distillable entanglement can be defined as follows.

\textbf{Definition 1}.-  The distillable total coherence $C_{R}^{D}(\rho)$ is defined by the optimal asymptotic transformation rate as
\begin{equation}
C_{R}^{D}(\rho)=R(\rho\rightarrow\left\vert\psi\right\rangle\left\langle\psi\right\vert)=\sup_{P}\lim_{n\rightarrow \infty}\frac{m}{n},\label{distill}
\end{equation}
where the supremum is optimized among all the potential transformation protocols $P$  with respect to the accuracy condition Eq. (\ref{condi}) and   $R(\rho\rightarrow\left\vert\psi\right\rangle\left\langle\psi\right\vert)$
denotes the optimal transformation rate.

(\textit{Total coherence cost}.-)The converse process can also dually and similarly be formulated as
which has been shown in Ref. \cite{horo} that
\begin{equation}
\left\vert\psi\right\rangle\left\langle\psi\right\vert^{\otimes m}\overset{\$_I}{\rightarrow}\sigma_m(\rho)\approx\rho^{\otimes n}
\end{equation}
such that
\begin{equation}
\lim_{n\rightarrow \infty}\left\Vert \sigma_m(\rho)-\rho^{\otimes n}\right\Vert_{tr}=0.
\end{equation}
 This process characterizes how much unit coherence is spent to prepare the given state, so the total coherence cost $C_R^C(\rho)$, similar to the entanglement cost, can be well defined.

 \textbf{Definition 2}.-The total coherence cost $C_R^C(\rho)$ is defined by the optimal asymptotic transformation rate $R(\left\vert\psi\right\rangle\left\langle\psi\right\vert\rightarrow\rho)$ as
\begin{equation}
C_R^C(\rho)=R(\left\vert\psi\right\rangle\left\langle\psi\right\vert\rightarrow\rho)=\inf_{P}\lim_{n\rightarrow \infty}\frac{m}{n},
\end{equation}

With the above definitions, we can present our second theorem.

\textbf{Theorem 2}.- For a state $\rho$, both the distillable total coherence  and the total coherence cost are the total coherence of $\rho$, that is,
\begin{equation}
C_R^C(\rho)=C_R^D(\rho)=C_R(\rho).
\end{equation}

\textbf{Proof}. This is actually an obvious conclusion obtained from Ref. \cite{horo} where the authors show that a pure state of qubit $\left\vert\psi\right\rangle$ can be asymptotically and reversibly transformed to a $d$
-dimensional state $\rho$ with the optimal transformation rate $R(\rho\rightleftharpoons\left\vert\psi\right\rangle\left\langle\psi\right\vert)=\log{d}-S(\rho)$ which is just the total coherence $C_R(\rho)$ given by Eq.
(\ref{1}).\hfill$\blacksquare$

Finally, we would like to emphasize that, since the total coherence is additive as given in Property (5), both the distillable total coherence and the total coherence cost are additive, i.e.,
$C_R^K(\rho\otimes\sigma)=C_R^K(\rho)+ C_R^K(\sigma)$ for any two states $\rho$ and $\sigma$ with $K$ representing $C$ and  $D$.

In addition, one could consider the similarities between the total coherence of formation and the entanglement of formation. We have
to emphasize that the total coherence is different from entanglement as well as the basis-dependent coherence. Suppose we formally define the total coherence of formation $C_R^f(\rho)$ for a $d$ -dimensional state $\rho$ by
the minimal average total coherence among all potential decomposition of a density matrix  as $\rho=\sum_k p_k\left\vert \phi_k\right\rangle\left\langle\phi_k\right\vert$. However, one will find that $C_R^f(\rho)=\log{d}$
since all the pure states has the same total coherence $\log{d}$. In addition, if one extends the pure-state decomposition to any form of decomposition (it could be meaningless), one can find the minimal total coherence is
just the total coherence of $\rho$, i.e., $C_R(\rho)$ due to the convexity of $C_R(\rho)$.

\subsection*{Converting total coherence to total correlation}
The resource theory as mentioned above essentially shows some quantum features as a resource, on the one hand, can be concentrated or diluted, and on the other hand, is implied to be spent on realizing some particular
 tasks or being converted to other forms. For example, quantum discord was shown to be converted to the distillable entanglement with the equal amount. In particular, recently it has been found that both quantum
 entanglement \cite{strel} can be obtained by the equal amount of basis-dependent coherence by the incoherent operations assisted by the auxiliary incoherent states. And quantum discord \cite{vedral} can also (may
 unequally)  be  obtained by the similar process. However, in what follows we will show that the total coherence can also be converted to the equal amount of total correlation by our defined incoherent operations with the
 assistance of the unique incoherent state.

To proceed, let's first introduce the total correlation in terms of the relative entropy. Let $\mathcal{P}$ denotes the set of all the product states, then the total correlation of a $(d_1\otimes d_2)$ -dimensional quantum
state $\rho_{AB}$ is defined by the distance between the state $\rho_{AB}$ and the set $\mathcal{P}$. According to the relative entropy, the total correlation denoted by $I(\rho_{AB})$ is actually the mutual information of
$\rho_{AB}$, that is,
\begin{eqnarray}
I(\rho_{AB})&=&\min_{\sigma\in\mathcal{P}}S\left(\rho||\sigma\right)\label{mutual0}\\
&=&S(\rho_A)+S(\rho_B)-S(\rho_{AB}),\label{mutual1}
\end{eqnarray}
 where
$S\left(\rho||\sigma\right)=\mathrm{Tr}\left\{\rho\log{\rho}-\rho\log{\sigma}\right\}$ denotes the relative entropy and $S(\cdot)$ is the von Neumann entropy.
Thus, we can present our theorems as follows.

\textbf{Theorem 3}.- Given an $n$ -dimensional state $\rho^S$, it can be converted to a composite state $\rho^{SA}=\$_I\left[ \rho^S\otimes \frac{\mathbb{I}^A_{m}%
}{m}\right]$ by  the incoherent operation $\$_I$ given in Property (2) with the assistance of the incoherent state $\frac{\mathbb{I}^A_m}{m}$. But the total correlation of $\rho^{SA}$ is restricted by the total coherence of
$\rho^S$ as
\begin{equation}
I\left( \$_I\left[ \rho^S\otimes \frac{\mathbb{I}_{m}%
}{m}\right] \right) \leq C_R\left( \rho^S \right).  \label{1re}
\end{equation}
In particular, there exists a unitary transformation $\$_I^o$ such that the upper bound can be achieved.

\textbf{Proof}. Based on the definition of the total coherence given in Eq. (\ref{1}), we can write
\begin{eqnarray}
C_R(\rho^S)&=&S\left(\rho^S\Vert\frac{\mathbb{I}_n}{n}\right)\notag\\
&=&S\left(\rho^S\otimes\frac{\mathbb{I}_m}{m}\Vert\frac{\mathbb{I}_n}{n}\otimes\frac{\mathbb{I}_m}{m}\right)\notag\\
&\geq&S\left(\$_I\left[\rho^S\otimes\frac{\mathbb{I}_m}{m}\right]\Vert\$_I\left[\frac{\mathbb{I}_n}{n}\otimes\frac{\mathbb{I}_m}{m}\right]\right)\label{dd1}\\
&\geq&\min_{\sigma\in\mathcal{P}}S\left( \$_I\left[ \rho^S\otimes \frac{\mathbb{I}_{m}%
}{m}\right]\Vert\sigma\right)\label{dd2}\\
&=&I\left( \$_I\left[ \rho^S\otimes \frac{\mathbb{I}_{m}%
}{m}\right] \right),\label{dd3}\end{eqnarray}
where Eq. (\ref{dd1}) comes from the contractibility of the relative entropy, Eq. (\ref{dd2}) is derived from the fact $\$_I\left[\frac{\mathbb{I}_n}{n}\otimes\frac{\mathbb{I}_m}{m}\right]=\frac{\mathbb{I}_{nm}}{nm}\in
\mathcal{P}$ and Eq. (\ref{dd3}) is due to the definition of the total correlation  (i.e., Eqs. (\ref{mutual0}) and (\ref{mutual1})). So Eq. (\ref{1re}) is satisfied. In particular, we have shown that inequality in Eq. (\ref{1re}) can be saturated in \textbf{Appendix}. This completes the proof. \hfill$\blacksquare$

\section{Relations with other quantum features}

The above the resource theory has provided a direct operational meaning for the total coherence, that is, the total coherence serves as the reversibly transformational rate between the resource state and the target state. In fact, the resource theory as a quantification of quantum feature also provides a way for us to understand one kind of quantum feature through another one. The most remarkable example is the monogamy of entanglement in a pure state of three qubits $\left\vert\psi\right\rangle_{ABC}$ which shows that 3-tangle $\tau \left(\left\vert\psi\right\rangle_{ABC}\right)$ can be understood by the residual entanglement as (r1) \cite{wootter1} $\tau \left(\left\vert\psi\right\rangle_{ABC}\right)=C^2\left(\rho_{A(BC)}\right)-C^2\left(\rho_{AB}\right)-C^2\left(\rho_{AC}\right)$ or (r2) \cite{yu3}  $\tau \left(\left\vert\psi\right\rangle_{ABC}\right)=C^2_a\left(\rho_{AB}\right)-C^2\left(\rho_{AB}\right)$ where  $\rho_{ABC}=\left\vert\psi\right\rangle_{ABC}\left\langle\psi\right\vert$, $\rho_{AB}$ and $\rho_{AC}$ denote the reduced density matrices and $C\left(\cdot\right)$ represents the bipartite concurrence of qubits and  $C_a\left(\rho_{AB}\right)$ denotes the localized (assisted) coherence defined by the average concurrence maximizing among all potential pure-state decomposition of $\rho_{AB}$. So Ref. \cite{yu2} considered the similar question for basis-dependent coherence and found that the similar understanding of the basis-dependent coherence only holds for $(2\otimes d)$ -dimensional system based on the $l_1$-norm coherence measure. The similar relations of the types (r1) and (r2) could be obtained by the $l_2$-norm coherence measure for any bipartite pure state, but $l_2$ norm is not a good coherence measure. Here we will show that the total coherence can be understood by the relations of both the types  (r1) and (r2).

At first, we would like to say that the total coherence inherited all the properties of the von Neumann entropy,   since it can be directly written by the von Neumann entropy, so it could have many interesting
implications. Consider the sub-additivity of von Neumann entropy, one can easily find that $C_R(\rho_{AB})\geq C_R(\rho_{A})+C_R(\rho_{B})$ for any bipartite state $\rho_{AB}$ with its reduced density matrices $\rho_{A/B}$. Thus we can see that
\begin{equation}
I(\rho_{AB})=C_R(\rho_{AB})-C_R(\rho_{A})-C_R(\rho_{B}).\label{tto}
\end{equation}
It is obvious that the total correlation $I(\rho_{AB})$ is just the residual total coherence between the nonlocal total coherence of a bipartite state and the two local coherences, which has just the similar form of type (r1). In particular, Eq. (\ref{tto}) holds for both pure and mixed state $\rho_{AB}$.

In addition, according to the strong sub-additivity of the von Neumann entropy, one can find that $C_R(\rho_{ABC})\geq C_R(\rho_{AB})+C_R(\rho_{BC})-C_R(\rho_B)$ for a tripartite state $\rho_{ABC}$ and the reduced density matrices $\rho_{AB}$,
$\rho_{BC}$ and $\rho_{B}$. These could provide the important reference for the distribution of coherence among multiple subsystems.

Now, let's consider a simple scheme between Alice and Bob who share a pure state $\left\vert\psi\right\rangle_{AB}$ of two qudits. Suppose Bob is allowed to perform any local operation on his qudit, the aim is to evaluate the average total coherence of Alice's qudit assisted by the classical communication with Bob. Since any operation could be performed by Bob, this is equivalent to that at Alice's side, she can obtain all the potential decomposition of her density matrix $\rho_A=\mathrm{Tr}_B\left\{\left\vert\psi\right\rangle_{AB}\left\langle\psi\right\vert\right\}$. Therefore, a trivial case is that Bob does nothing or a trivial identity operation. Alice's total coherence is just the total coherence of the reduced density matrix $\rho_A$, i.e., $C_R(\rho_A)=\log{n}-S(\rho_A)$ with $n$ denoting the dimension of her qudit. Due to the convexity of the total coherence, in this case Alice obtains the minimal average total coherence. On the contrary, Bob can apply a general positive-operator-valued-measure on his side such that Alice always obtains a pure state with some probability. Then Alice's average total coherence will reach the maximal value $C_R^a(\rho_A)=\log{n}$ with the superscript $a$ representing the maximal value. Thus the discrepancy between Alice's maximal and minimal total coherence is give by
\begin{equation}
E(\left\vert\psi\right\rangle_{AB})=C_R^a(\rho_A)-C_R(\rho_A)\label{tto2}
\end{equation}
with $E(\left\vert\psi\right\rangle_{AB})=S(\rho_A)$ being the von-Neumann-entropy entanglement of $\left\vert\psi\right\rangle_{AB}$. It is obvious that Eq. (\ref{tto2}) has the same form as (r2).
In addition, one can easily find that for a mixed state $\rho_{AB}$, Eq. (\ref{tto2}) can lead to
\begin{equation}
E(\rho_{AB})\leq C_R^a(\rho_A)-C_R(\rho_A).\label{tto3}
\end{equation}
where $E(\rho_{AB})=\min_{\{p_i,\left\vert\psi_i\right\rangle\}}\sum_ip_iE(\left\vert\psi_i\right\rangle)$ with $\rho_{AB}=\sum_ip_i\left\vert\psi_i\right\rangle\left\langle\psi_i\right\vert$ is the mixed-state entanglement. This can be easily proven  as follows.  Let's consider a purification state $\left\vert\psi\right\rangle_{ABC}$  such that $\rho_{AB}=\mathrm{Tr}_C\left\vert\psi\right\rangle_{ABC}\left\langle\psi\right\vert$, based on Eq. (\ref{tto2}), we have $E(\left\vert\psi\right\rangle_{A(BC)})=C_R^a(\rho_A)-C_R(\rho_A)$. Since the entanglement isn't increased by the local operations, tracing over subsystem $C$ will decrease the entanglement, i.e., $E(\left\vert\psi\right\rangle_{AB})\leq E(\left\vert\psi\right\rangle_{A(BC)})$ which directly leads to Eq. (\ref{tto3}).

\section{Discussion}

Since it has been shown that entanglement and quantum discord can be obtained from the basis-dependent coherence by the corresponding incoherent operation. One could wonder whether the total correlation could be converted
to the entanglement (or quantum discord) by the incoherent operation assisted by the incoherent state $\frac{\mathbb{I}_N}{N}$. In fact, one can find that there exist some incoherent operations to realize the conversion,
but in general the total coherence could not be completely converted to entanglement. In other words, some total coherence could be converted to the classical correlation, so that the total coherence can be completely converted to the
total correlation instead of the quantum correlation. A simple example can demonstrate this. Suppose we employ a unitary transformation to convert a pure state to a composite entangled state with the assistance of a maximally mixed state
$\frac{\mathbb{I}_N}{N}$. As we know, the pure states have the maximal total coherence, so one could expect to obtain a maximally entangled state through the above conversion.  However, one can note that the pre-conversion
state is a mixed product state, but the fact is that a unitary transformation will never transform a mixed state to a pure state which possesses the maximal entanglement. One could intend to employ the general incoherent
operation $\$_I$, however, the convexity of a valid entanglement measure implies that $\$_I$ will lead to the less entanglement than the unitary operation. Thus this shows that the maximal total coherence cannot be
completely converted to the entanglement.

However, from a general application point of view, i.e., beyond the resource theory restricted by our Properties (1-5), it is trivial to find that the nonzero total coherence can be completely converted to the entanglement
or quantum discord. This is implied in the definition of the total coherence. That is, one can always find a suitable reference framework (basis) to make the considered state has the equal amount of coherence with respect
to the reference framework. So in this framework, one can employ the scheme in Ref. \cite{strel} to convert the coherence to the equal amount of entanglement.
\section{Conclusion}
We have established the operational resource theory for the total coherence by
showing the reversible transformation between the resource state (pure state) and any given state by the defined incoherent operations both in the single-copy regime and the asymptotic regime. In addition, we extensively
find that the total coherence can be converted to the equal amount of total correlation by the incoherent operations assisted by the maximally mixed state. Finally we find that the total correlation of a bipartite state can be understood by the residual total coherence between the nonlocal total coherence of a bipartite state and the two local total coherence, which is also an operational meaning of the total coherence. In addition, we also show that the pure-state entanglement can be understood by the discrepancy of the maximal and the minimal local total coherence, which is another interesting understanding of the total coherence.

\section*{Acknowledgement}
This work was supported by the National Natural Science
Foundation of China, under Grant No.11775040 and 11375036, and the Xinghai Scholar
Cultivation Plan.

\noindent \textit{Noted added}.---When we finished this paper, we noticed Ref. \cite{similar} had also outcomes similar to some of our results. 
\section{Appendix}

\textbf{Proof of the equality of theorem 3:} In the proof of theorem 3, we have shown that the inequality is satisfied. Here we will prove the inequality in Eq. (\ref{1re}) can be saturated. Let's consider  the state $\rho^S$ again. The total coherence of $\rho^S$ is a basis-free coherence measure and quantify the maximal (basis-dependent)
coherence with all potential basis taken into account. So one can also select an incoherent operation $\tilde{U}$ (i.e., the unitary transformation) to transform $\rho^S$ into
$\tilde{\rho}^S=\tilde{U}\rho^S \tilde{U}^\dag=\sum_{ij}\tilde{\rho}_{ij}\left\vert i\right\rangle\left\langle j\right\vert$
with $\tilde{\rho}_{ij}$ denoting the matrix entries in the $\{\left\vert i\right\rangle\}$  representation,  the basis-dependent coherence in this representation is just equal to the total coherence. It is obvious that in
this representation the diagonal entries of $\tilde{\rho}^S$ is uniform, i.e., $\tilde{\rho}_{ii}=\frac{1}{n}$.
Thus one can consider another incoherent operation on a composite system (denoted, respectively, by $S$ and $A$) defined as $\Lambda_{I}\left[ \rho _{SA}%
\right] =U\rho _{SA}U^{\dagger }$ with
\begin{equation}
U=\sum\limits_{i=0}^{n-1}\sum\limits_{j=0}^{m-1}\left\vert
i\right\rangle \left\langle i\right\vert _{S}\otimes |\mathrm{mod}\left(
i+j,m\right) \rangle \left\langle j\right\vert _{A}
\end{equation}
This unitary operation $U$ is actually the generalized Controlled-NOT gate in a high dimensional system. With this unitary operation, for the state $\tilde{\rho}^S$ and any orthonormal basis $\{\left\vert k\right\rangle\}$
in a $m$ -dimensional subsystem $A$, one can find that
\begin{eqnarray}
&&\Lambda _{I}\left[ \tilde{\rho} ^{S}\otimes\frac{\mathbb{I}_m^A}{m}\right]=\frac{1}{m}\sum_k \Lambda _{I}\left[ \tilde{\rho} ^{S}\otimes  \left\vert k\right\rangle
\left\langle k\right\vert _{A}\right] \notag \\
&=&\sum_k^{m-1}\sum\limits_{i_{1},i_{2}=0}^{n-1}\frac{\tilde{\rho} _{i_{1}i_{2}}}{m}\left\vert
i_{1}\right\rangle \left\langle i_{2}\right\vert _{S}\otimes \left\vert
i_{1}\oplus k\right\rangle_{A} \left\langle i_{2}\oplus k\right\vert.\label{sss}
\end{eqnarray}

With the unitary operation Controlled-NOT gate, for the state $\tilde{\rho}^S$ and any orthonormal basis $\{\left\vert k\right\rangle\}$
in a $m$ -dimensional subsystem $A$, we have
\begin{eqnarray}
&&\Lambda _{I}\left[ \tilde{\rho} ^{S}\otimes\frac{\mathbb{I}_m^A}{m}\right]=\frac{1}{m}\sum_k \Lambda _{I}\left[ \tilde{\rho} ^{S}\otimes  \left\vert k\right\rangle
\left\langle k\right\vert _{A}\right] =\sum_k U\frac{\tilde{\rho} ^{S}}{m}\otimes \left\vert k\right\rangle _{A}\left\langle k\right\vert
U^{\dagger }\notag \\
&=&\sum_{k=0}^{m-1}\left(\sum\limits_{i=0}^{n-1}\sum_{j=0}^{m-1}\left\vert
i\right\rangle_{S} \left\langle i\right\vert \otimes \left\vert \mathrm{mod}\left(i+j,m\right)\right\rangle _{A}\left\langle j\right\vert
\right) \\
&&\sum\nolimits_{i,j}\frac{\tilde{\rho} _{ij}}{m}\left\vert i\right\rangle_{S} \left\langle
j\right\vert \otimes \left\vert k\right\rangle_{A} \left\langle k\right\vert
\left(\sum\limits_{i=0}^{n-1}\sum_{j=0}^{m-1}\left\vert
i\right\rangle_{S} \left\langle i\right\vert \otimes \left\vert
j\right\rangle_{A} \left\langle \mathrm{mod}\left( i+j,m\right) \right\vert  \right)\notag\\
&=&\sum_k^{m-1}\sum\limits_{i_{1}=0}^{n-1}\sum\limits_{i_{2}=0}^{n-1}\left(
\left\vert i_{1}\right\rangle_{S} \left\langle i_{1}\right\vert
\sum\nolimits_{i,j}\frac{\tilde{\rho} _{ij}}{m}\left\vert i\right\rangle _{S} \left\langle
j\right\vert\cdot\left\vert i_{2}\right\rangle _{S}\left\langle i_{2}\right\vert
\right) \notag\\
&&\otimes \left( \sum\limits_{j_{1}=0}^{m-1}\left\vert \mathrm{mod}\left(
i_{1}+j_{1},m\right) \right\rangle_{A} \left\langle j_{1}\right\vert\cdot
\left\vert k\right\rangle_{A} \left\langle k\right\vert
\right.\left.\sum\limits_{j_{2}=0}^{m-1}\left\vert j_{2}\right\rangle _{A}
\left\langle \mathrm{mod}\left( i_{2}+j_{2},m\right) \right\vert\right)  \notag\\
&=&\sum_k^{m-1}\sum\limits_{i_{1},i_{2}=0}^{n-1}\frac{\tilde{\rho} _{i_{1}i_{2}}}{m}\left\vert
i_{1}\right\rangle \left\langle i_{2}\right\vert _{S}\otimes \left\vert
i_{1}\oplus k\right\rangle_{A} \left\langle i_{2}\oplus k\right\vert.
\end{eqnarray}

From Eq. (\ref{sss}), one can find that the  reduced matrices of $\Lambda _{I}\left[ \tilde{\rho} ^{S}\otimes\frac{\mathbb{I}_m^A}{m}\right]$ can be given by
\begin{eqnarray}
\left\{
\begin{array}{c}\sigma_{S}=\mathrm{Tr}_{A} \Lambda _{I}\left[ \tilde{\rho} ^{S}\otimes\frac{\mathbb{I}_m^A}{m}\right]=\frac{\mathbb{I}_{n}}{n}, \\\sigma_{A}=\mathrm{Tr}_{S} \Lambda _{I}\left[ \tilde{\rho}
^{S}\otimes\frac{\mathbb{I}_m^A}{m}\right]=\frac{\mathbb{I}_{m}}{m},\end{array}\right.\label{redusss}
\end{eqnarray}
where we consider the fact that we have made $\tilde{\rho}_{ii}=\frac{1}{n}$ for all $i$'s.
Thus the total correlation (i.e., the mutual information) of $\Lambda _{I}\left[ \tilde{\rho} ^{S}\otimes\frac{\mathbb{I}_m^A}{m}\right]$ can be calculated as
\begin{eqnarray}
&&I\left( \Lambda _{I}\left[ \tilde{\rho} ^{S}\otimes\frac{\mathbb{I}_m^A}{m}\right]\right)\notag\\
&=&S\left(\sigma_S\right)+S\left(\sigma_A\right)-S\left(\Lambda _{I}\left[ \tilde{\rho} ^{S}\otimes\frac{\mathbb{I}_m^A}{m}\right]\right)\notag\\
&=&\log{n}+\log{m}-S\left( \tilde{\rho} ^{S}\otimes \frac{\mathbb{I}_m^A}{m}\right) \label{zhongy0}\\
&=&\log{n}-S\left( \tilde{\rho} ^{S}\right)=C_R\left(\rho ^{S}\right).\label{zhongy1}
\end{eqnarray}
Eq. (\ref{zhongy0}) holds because the von Neumann entropy $S(\rho^S)$ doesn't change under the unitary transformation $\tilde{\rho}^S=\tilde{U}\rho^S\tilde{U}^\dag$ and Eq. (\ref{zhongy1}) holds due to the additivity of von
Neumann entropy for product states.
Thus we show that there exists a unitary transformation $U_T=U\tilde{U}$ such that Eq. (\ref{1re}) is satisfied, which completes the proof. \hfill$\blacksquare$

\end{document}